# Photoluminescence from Single-Walled MoS$_2$ Nanotubes Coaxially Grown on Boron Nitride Nanotubes

Ming Liu[1], Kaoru Hisama[1], Yongjia Zheng[1], Mina Maruyama[2], Seungju Seo[1], Anton Anisimov[3], Taiki Inoue[1,4], Esko I. Kauppinen[5], Susumu Okada[2], Shohei Chiashi[1], Rong Xiang[1], and Shigeo Maruyama[1]*

[1] *Department of Mechanical Engineering, The University of Tokyo, Tokyo 113-8656, Japan*

[2] *Graduate School of Pure and Applied Sciences, University of Tsukuba, Tsukuba 305-8571, Japan*

[3] *Canatu Ltd., Helsinki FI‐00390, Finland*

[4] *Department of Applied Physics, Graduate School of Engineering, Osaka University, Osaka 565-0871, Japan*

[5] *Department of Applied Physics, Aalto University School of Science, Espoo 15100, FI-00076 Aalto, Finland*

*email: maruyama@photon.t.u-tokyo.ac.jp

**Abstract**

Single- and multi-walled molybdenum disulfide ($MoS_2$) nanotubes have been coaxially grown on small diameter boron nitride nanotubes (BNNTs) which were synthesized from heteronanotubes by removing single-walled carbon nanotubes (SWCNTs), and systematically investigated by optical spectroscopy. The strong photoluminescence (PL) from single-walled $MoS_2$ nanotubes supported by core BNNTs is observed in this work, which evidences a direct band gap structure for single-walled $MoS_2$ nanotubes with around 6 - 7 nm in diameter. The observation is consistent with our DFT results that the single-walled $MoS_2$ nanotube changes from an indirect-gap to a direct-gap semiconductor when the diameter of a nanotube is more than around 5 nm. On the other hand, when there are SWCNTs inside the heteronanotubes of BNNTs and $MoS_2$ nanotubes, the PL signal is considerably quenched. The charge transfer and energy transfer between SWCNTs and single-walled $MoS_2$ nanotubes were examined through characterizations by PL, XPS, and Raman spectroscopy. Unlike the single-walled $MoS_2$ nanotubes, multi-walled $MoS_2$ nanotubes do not emit light. Single- and multi-walled $MoS_2$ nanotubes exhibit different Raman features in both resonant and non-resonant Raman spectra. The method of assembling heteronanotubes using BNNTs as templates provides an efficient approach for exploring the electronic and optical properties of other transition metal dichalcogenide nanotubes.

**Introduction**

Van der Waals heterostructure of two-dimensional (2D) layered materials[1] have attracted much attention in material research since the first experimental isolation of single-layer graphene.[2-7] The novel properties in heterostructures with diverse layerings of metal, semiconductors or insulators have raised numerous new designs of electronic devices as well as optoelectronic devices. Atomically thin transition-metal dichalcogenide (TMD) such as $MoS_2$ holds great promise for electrical, optical, and mechanical devices, as well as novel physical phenomena.[8] Meanwhile, wrapping 2D TMD materials into 1D nanotubes has brought interests in creating chiral tubular structure and radial heterojunctions with diverse functionalities.[9] Recently, we have demonstrated one-dimensional (1D) van der Waals heterostructures templated on single-walled carbon nanotubes (SWCNTs).[10] 1D van der Waals materials will join the innovation of electronic and optoelectronic devices with 2D layered materials.

Fullerene-like and nanotube structures of $MoS_2$ have been produced by a gas-phase reaction firstly in 1995.[11] A combined theoretical and experimental work has indicated that multi-walled $MoS_2$ nanotubes are more stable than single-walled $MoS_2$ nanotubes, while single-walled $MoS_2$ with diameter more than about 6 nm is more stable than nanoribbon shape.[12] The conventional growth process has resulted in multi-walled $MoS_2$ nanotubes with large diameters.[11, 13] The bulk form of $MoS_2$ is an indirect band gap semiconductor with an energy gap of ~1.2 eV[14] and has attracted attention as photovoltaic and photocatalytic materials.[15, 16] The band gap of $MoS_2$ changes from indirect to direct when the thickness reduces to a single layer.[17] In addition, recent studies have showed the strong photoluminescence in monolayer $MoS_2$ that can be attributed to the direct gap electronic structure.[18-20] The band structure of single-walled and multi-walled $MoS_2$ nanotubes are still not well investigated experimentally, though modulated calculations have predicted that $MoS_2$ nanotubes have indirect band gaps except those with zigzag chiral indices.[21-23] Here, although the zigzag $MoS_2$ nanotubes are calculated as direct-gap semiconductors, the band edge states are predicted to come from different subbands and PL is not expected.[24]

Using a tubular template to confine the formation of inorganic nanotubes has been proposed to stabilize few-walled inorganic nanotubes, which is a critical step toward understanding their optical properties and implementing into novel devices. Carbon nanotubes (CNTs) have been used as a template to assemble specific materials[25, 26] and as filling vessels for metals[27, 28] for forming core-shell nanotubes. Charge transfer from a CNT core to a $MoS_2$ sheath has been verified by X-ray photoelectron spectroscopy in a composite $MoS_2$/CNT material.[29] The ultrafast optoelectronic processes in 1D heterostructures have been investigated lately.[30] Core-shell carbon@$MoS_2$ nanotube sponges have been utilized as electrodes in a high-performance battery.[31] However, single-walled $MoS_2$ nanotubes are not observed in the previous core-shell inorganic nanotube studies. Besides, CNTs are optically active materials with high response signal of Raman spectra[32] and photoluminescence spectra,[33] which makes the investigation of optical properties of the TMD shell outside a CNT infeasible. Hence, boron nitride nanotube (BNNT) is a promising substitute for serving as a template for forming $MoS_2$ nanotubes since the band gap of BNNTs is ~5.5 eV which is large enough to be transparent within a wide range of wavelengths and quasi-independent from the chirality.[34]

In this paper, we present a systematic study of the optical properties and electronic structure of $MoS_2$ nanotubes including single-walled and multi-walled types. We have developed a facile chemical vapor deposition (CVD) method for synthesizing core-shell boron nitride and $MoS_2$ heteronanotubes (BNNT@$MoS_2$NT) by removing the SWCNTs in heteronanotubes of SWCNTs coated by BNNT (SWCNT@BNNT). The hetero structure based on BNNTs is a promising platform for studying the optical properties of $MoS_2$ layers or other TMD layers. The properties of BNNT@$MoS_2$NT heteronanotubes were examined by using three complementary spectroscopic techniques: optical absorption, Raman spectroscopy, and photoluminescence (PL), with additional characterizations provided by transmission electron microscopy (TEM) and DFT calculations using the STATE (Simulation Tool for Atom Technology) package. The strong PL signal from single-walled $MoS_2$ nanotubes indicates the direct band gap structure of the material, which is consistent with our DFT results. With increasing the wall number of $MoS_2$ nanotubes, the PL signal decreased significantly which allows us to trace the direct and indirect band gaps structures of the $MoS_2$ nanotubes. The charge and energy transfer between SWCNTs and $MoS_2$ nanotubes in the 1D hetero structure were observed in this work.

## Results and Discussion

**Synthesis and structure analysis of BNNT@MoS$_2$NT heteronanotubes.** In this work, we presented a template-assisted approach involving a sequence of facile CVD processes using SWCNTs as a template to produce BNNT@MoS$_2$NT heteronanotubes. The SWCNT film prepared by areosal CVD was used as a template and transferred onto a ceramic washer with the size of 4 mm inner diameter and 6 mm outer diameter. Firstly, the suspended SWCNT film on the washer was used as a template and ammonia borane was employed as the precursor to form a coaxial tube structure of SWCNT coated by boron nitride nanotubes (SWCNT@BNNT). Then the SWCNT@BNNT film was annealed at 610°C in the oxygen atmosphere with the pressure of 85 kPa for 12 hours to remove the SWCNTs in the film. The optical image of the as-synthesized BNNT film after the annealing process is shown in Figure 1a. The BNNT film is fully transparent within the visible range of wavelengths, which is confirmed by the absorption spectrum in Figure S1a. The Raman peak at ∼1370 cm$^{-1}$ was detected in Figure S1b, which is corresponding to the $E_{2g}$ in-plane vibrational mode of the h-BN networks. The Raman peak at 1555.5 cm$^{-1}$ in Figure S1b is the vibrational mode of oxygen. In the Fourier transformed infrared (FT-IR) spectrum (Figure S1c), the absorption band at ∼1369.5 cm$^{-1}$ is attributed to the in-plane stretching mode of the h-BN that vibrates along the longitudinal or tube axis of a BNNT. Besides, the peak at ~1520 cm$^{-1}$ corresponds to the stretching of the h-BN along the tangential directions of a BNNT. This stretching mode smears out for h-BN bulks or thin films, and only shows up when the tube curvature induces an anisotropic strain on the h-BN which indicates the high crystallinity of BNNTs.[35] The representative high-resolution TEM image and the atomic model of a high quality BNNT are presented in Figure 1, b and c. The as-synthesized BNNT film on a ceramic was a template for the MoS$_2$ CVD process to form the BNNT@MoS$_2$NT heteronanotubes. The MoS$_2$ CVD was conducted at 530 °C with the sulfide and MoO$_3$ powders as precursors. The typical growing time of MoS$_2$ at 530 °C is 10 min and the optical image of BNNT@MoS$_2$NT film is shown in Figure 1d. In Figure 1e, the representative TEM image of the BNNT@MoS$_2$NT, the diameter of the outer layer MoS$_2$ nanotube is around 6.8 nm and the inner diameter of the multi-walled BNNT is 2.1 nm. The atomic model of a BNNT@MoS$_2$NT heteronanotube is presented in Figure 1f. In the BNNT@MoS$_2$NT heteronanotube, there is no chemical bonding between BN layer and MoS$_2$ layer.

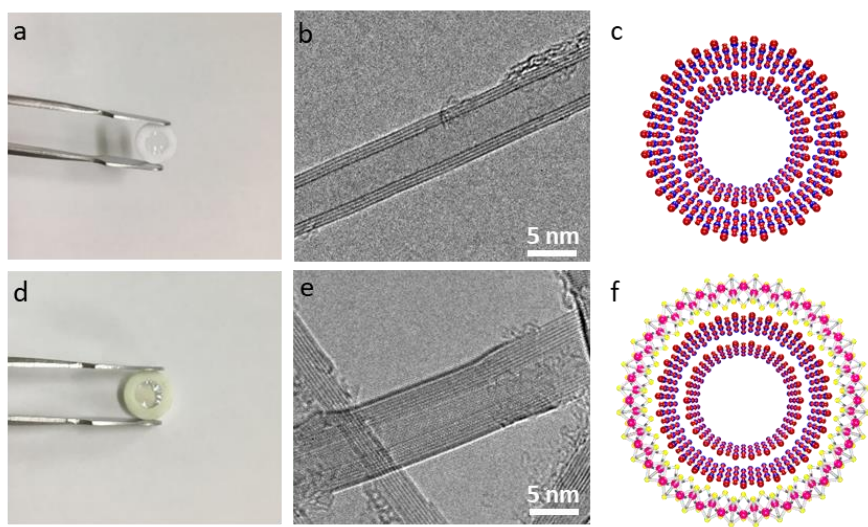

Figure 1. (a) Optical image of the suspended BNNT film on a ceramic washer. (b) Representative high-resolution TEM image of a multi-walled BNNT. (c) Atomic model of a multi-walled BNNT. (d) Optical image of the suspended BNNT@MoS$_2$NT nanotubes film on a ceramic washer. (e) Representative high-resolution TEM image of a BNNT@MoS$_2$NT. (f) Atomic model of a BNNT@MoS$_2$NT.

**Photoluminescence from single-walled MoS$_2$ nanotubes and the charge and energy transfer between SWCNT and MoS$_2$ nanotubes.** Figure 2a shows the atomic model of two types of heteronanotubes that we used to compare, SWCNT@BNNT@MoS$_2$NT and BNNT@MoS$_2$NT heteronanotubes. The PL, Raman, and absorption spectra of the two heteronanotubes are compared in Figure 3b, 3c, and 3d, respectively. The SWCNT@BNNT@MoS$_2$NT heteronanotube was synthesized by the serial CVD method described in our previous work.[10] The representative TEM image of the SWCNT@BNNT@MoS$_2$NT heteronanotube is shown in Figure S2. Optical measurements by PL, Raman, and absorption spectroscopy were performed on the free-standing SWCNT@BNNT@MoS$_2$NT and BNNT@MoS$_2$NT heteronanotube films on ceramic washers. All optical measurements were conducted under ambient conditions at room temperature. The excitation wavelength of Raman and PL spectroscopy was 532 nm (2.33 eV) with a spot diameter of ~2 μm focused on the sample. A low laser power of ~70 μW (out of the microscope) was applied to avoid heating and PL saturation.[36] Appreciable PL was observed from single-walled MoS$_2$ nanotubes coaxially grown on BNNTs, the BNNT@MoS$_2$NT samples. The measured PL intensities for a BNNT@MoS$_2$NT and a SWCNT@BNNT@MoS$_2$NT sample under the identical excitation at 2.33 eV are significantly different (Figure 2b). The PL intensity of

SWCNT@BNNT@MoS$_2$NT sample is eight times less than the PL signal from the BNNT@MoS$_2$NT sample. However, the shape of the normalized PL spectra for both samples are nearly identical. The PL spectrum of BNNT@MoS$_2$NT consists a single feature centered at 1.85 eV. In contrast, the center of the PL peak for SWCNT@BNNT@MoS$_2$NT is at 1.87 eV, slightly blue shifted. The sharp peaks in both PL spectra are from the Raman signal of the MoS$_2$ and SWCNT in the film.

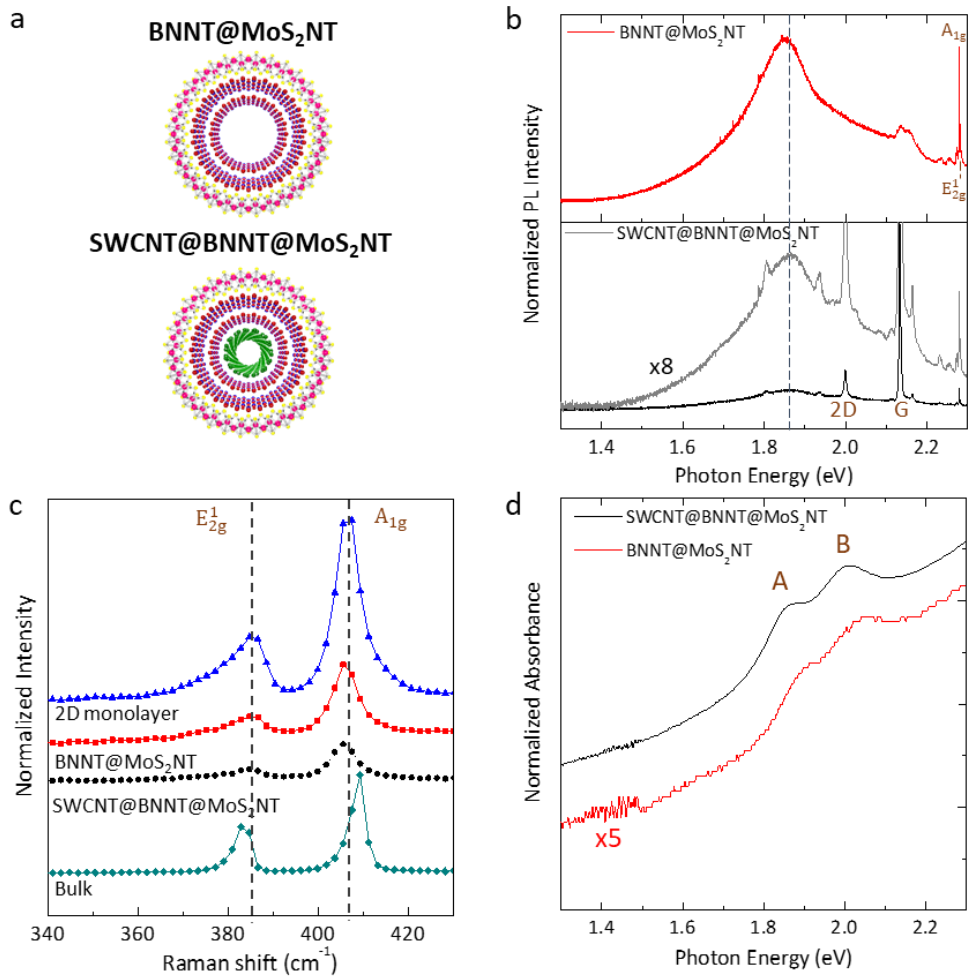

Figure 2. (a) The atomic model of a SWCNT@BNNT@MoS$_2$NT heteronanotube and a BNNT@MoS$_2$NT heteronanotube. (b) The PL spectra of BNNT@MoS$_2$NT and BNNT@MoS$_2$NT, respectively. The black dot line is the eight times enlarged PL spectrum of SWCNT@BNNT@MoS$_2$NT. (c) Representative Raman spectra of 2D monolayer MoS$_2$, BNNT@MoS$_2$NT, SWCNT@BNNT@MoS$_2$NT, and bulk MoS$_2$. (d) Representative absorption spectra of free-standing SWCNT@BNNT@MoS$_2$NT and BNNT@MoS$_2$NT films.

To understand the origin of the extraordinary PL property from single-walled MoS$_2$ nanotubes, we compared the PL spectra with the absorption spectra of these samples (Figure 2d). In the range of 1.3–2.3 eV, there are two main absorption peaks that are contributed to the direct-gap transitions between the maxima of split valence bands ($v1$ and $v2$) and the minimum of the conduction band ($c1$) which are all located at the K point of the Brillouin zone.[37-39] The PL peak of BNNT@MoS$_2$NT at 1.85 eV matches the lower absorption resonance (conventionally called A exciton) in the absorption spectrum. Therefore, we assigned the PL of BNNT@MoS$_2$NT to direct-gap luminescence. The upper absorption resonance (B exciton) is observed in SWCNT@BNNT@MoS$_2$NT and BNNT@MoS$_2$NT, however the PL emission from B exciton is not observed in our experiment.

The PL quenching in the SWCNT@BNNT@MoS$_2$NT sample is significant compared with the PL of BNNT@MoS$_2$NT. The near-field coupling, for example interlayer charge and/or energy transfer, is the main reason for causing the PL quenching effect.[40] To analyze the interaction between the SWCNT and MoS$_2$, we converted the PL spectrum of SWCNT@BNNT@MoS$_2$NT to the Raman scale and enlarged PL peak for both samples (Figure S3). In Figure S3a, there is a noticeable blue shift of the PL from SWCNT@BNNT@MoS$_2$NT compared with the PL from BNNT@MoS$_2$NT which verifies the photoinduced hole doping (p-dope) in MoS$_2$ nanotubes.[41, 42] The Raman spectrum of SWCNT in SWCNT@BNNT@MoS$_2$NT heteronanotubes (Figure S3b black line) presents the typical electron doping (n-dope) effect based on the downshift of the G and 2D peak position.[43, 44] In the XPS spectra (Figure S3c), the C1s peak of SWCNT@BNNT@MoS$_2$NT heteronanotubes presents 0.2 eV shifting to higher binding energy compared with the C1s peak of pristine SWCNT film which also proves the n-dope of SWCNT in the heteronanotubes. Therefore, in SWCNT@BNNT@MoS$_2$NT heteronanotubes, the net electron transfers from MoS$_2$ nanotubes to SWCNT causing the n-dope of SWCNT, and which naturally implies hole accumulated in MoS$_2$ resulting in the doping to the p-direction of MoS$_2$ nanotubes. Hence, charge transfer occurs between SWCNT and MoS$_2$ nanotubes in SWCNT@BNNT@MoS$_2$NT heteronanotubes. Although electrons may transfer to SWCNT referring to the demonstration from our experimental data, the interlayer charge transfer processes alone cannot be responsible for the massive PL quenching. Instead, interlayer energy transfer provides a highly efficient relaxation pathway[40] for excitons in SWCNT@BNNT@MoS$_2$NT heteronanotubes. Therefore, in SWCNT@BNNT@MoS$_2$NT heteronanotubes, charge transfer and energy transfer may happen at the same time. More detailed investigations need to be applied to confirm the hypothesis.

Raman spectra of 2D monolayer MoS$_2$, BNNT@MoS$_2$NT, SWCNT@BNNT@MoS$_2$NT, and bulk MoS$_2$ are shown in Figure 2c. The in-plane $E_{2g}^1$ and out-of-plane $A_{1g}$ Raman modes of MoS$_2$ were observed in our BNNT@MoS$_2$NT and SWCNT@BNNT@MoS$_2$NT samples. The frequencies of both modes in single-walled MoS$_2$ nanotubes in heteronanotubes are almost same as the monolayer MoS$_2$ values. The frequency difference of two modes is about 21 cm$^{-1}$ which is identical with the monolayer MoS$_2$ sample. From the previous study, the $E_{2g}^1$ vibration softens (red shifts), while the $A_{1g}$ vibration stiffens (blue shifts) with increasing thickness of 2D MoS$_2$ films.[45] However, we did not observe any clear softening or stiffening of the Raman modes in single-walled MoS$_2$ nanotubes compared with monolayer MoS$_2$ though the curvature and strain energy in nanotubes are different.

**Band structure of single-walled MoS$_2$ nanotubes.** A monolayer MoS$_2$ is well known as a semiconductor with a direct band gap at the K point in the Brillouin zone,[19] while MoS$_2$ nanotubes have been predicted as a semiconductor with an indirect band gap except zigzag types.[21] In the previous studies, the band edge is based on the effect of strain from the curvature, which induces an increase of the energy levels of valence bands around Γ point that forms an indirect band gap with the bottom of the conduction band originated from the monolayer K point. For the zigzag MoS$_2$ nanotubes, however, they have direct band gaps because the K and Γ points are folded at the same place in the one-dimensional Brillouin zone of a nanotube. Because this special folding is only possible for the zigzag nanotube, all chiral nanotubes are expected behave as armchair nanotubes.[22] Here we suspected that the effect of the strain would decrease as the diameter of a nanotube becomes larger. When the diameter of a nanotube reached a certain size, reduced strain would make the band gap of a nanotube approach the direct band gap as a monolayer MoS$_2$. Figure 3 presents the electronic structure of three types of single-walled MoS$_2$ nanotubes with chiral indices of (12,12), (24,24), and (36,36) that are calculated by a density functional theory.[46, 47] The energy is measured with respect to that of the valence band edge at the Γ point. The valence peak at near $k = 2\pi/3$, originated from the K point of the Brillouin zone, becomes higher when the diameter of the nanotube increases. For the chiral index of (36,36) with diameter about 6.3 nm, the band structure of a nanotube forms a direct bang gap as in Fig. 3c. The detailed discussion of the band gap crossover for MoS$_2$ nanotubes will be reported in another paper.[48] The result in our paper[48] shows when the diameter of a MoS$_2$ nanotube is larger than 5 nm, the nanotube is a semiconductor with a direct band gap which corresponds with the strong PL signal observed in the BNNT@MoS$_2$NT heteronanotube in this paper.

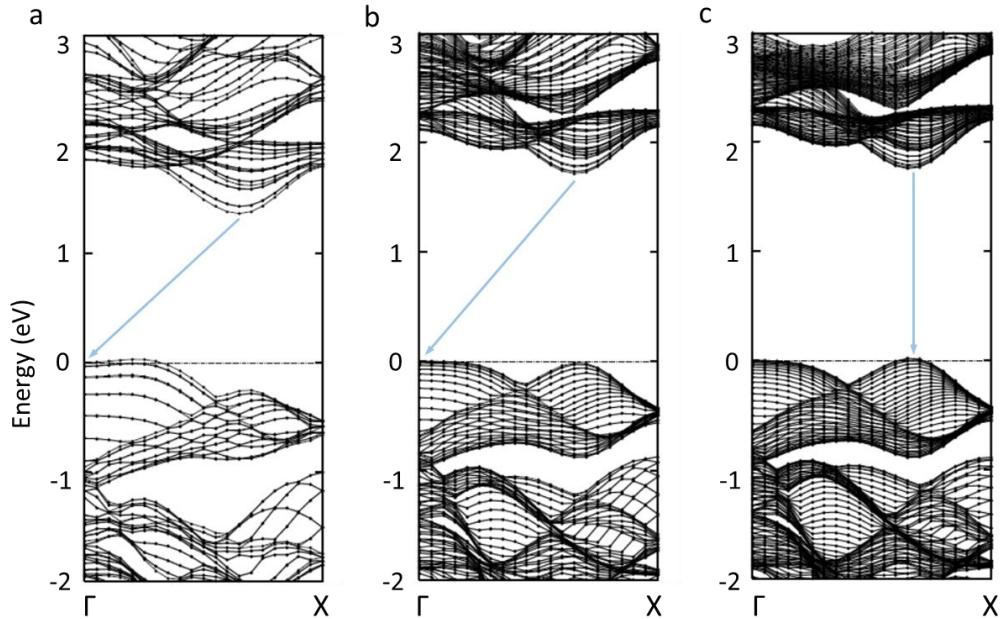

Figure 3. Calculated electronic structure of MoS$_2$ nanotubes with different indices (a) (12,12) with diameter of 2.1 nm, (b) (24,24) with diameter of 4.2 nm, and (c) (36,36) with diameter of 6.3 nm.

**Structure of BNNT@single-MoS$_2$NT and BNNT@multi-MoS$_2$NT heteronanotubes.** The heteronanotubes of single-walled MoS$_2$ nanotubes coaxially grown on BNNTs (BNNT@single-MoS$_2$NT) were systematically characterized by optical measurements for example PL, Raman spectroscopy, and absorption. Here, multi-walled MoS$_2$ nanotubes coaxially grown on BNNTs (BNNT@multi-MoS$_2$NT) were synthesized by modifying the MoS$_2$ CVD parameters for studying the difference of the optical properties between BNNT@single-MoS$_2$NT and BNNT@multi-MoS$_2$NT nanotubes. The representative TEM images of BNNT@single-MoS$_2$NT and BNNT@multi-MoS$_2$NT nanotubes are shown in Figure S4. In Figure S4a, the MoS$_2$ nanotubes coaxially grown on BNNTs are single-walled nanotubes, whereas the MoS$_2$ nanotubes in Figure S4b have two or three walls. These two types of samples were used in the following optical characterizations.

**The comparison of Raman, PL, and absorption spectra of BNNT@single-MoS₂NT and BNNT@multi-MoS₂NT heteronanotubes.** Figure 4a shows representative Raman spectra of BNNT@single-MoS₂NT and BNNT@multi-MoS₂NT samples with the excitation wavelength of 532 nm in air ambient environment. The $E_{2g}^1$ and $A_{1g}$ modes were observed in both BNNT@single- and BNNT@multi- MoS₂NT samples. The in-plane $E_{2g}^1$ mode is associated with opposite vibration of two S atoms with respect to the Mo atom, whereas the out-of-plane $A_{1g}$ mode origins from the vibration of S atoms in opposite directions.[49, 50] For BNNT@single-MoS₂NT sample, the peak of the $A_{1g}$ mode is at 405 cm⁻¹ and the $E_{2g}^1$ mode locates at 384 cm⁻¹. Strikingly, we found the $E_{2g}^1$ vibration softened to ~380 cm⁻¹ (red shifts), while the $A_{1g}$ mode remained at the same position (~405 cm⁻¹) in the BNNT@multi-MoS₂NT sample. When the wall number increases, the interlayer van der Waals force in MoS₂ suppresses atom vibration, resulting in higher force constants.[51] However, the softening of $E_{2g}^1$ vibration and the negligible change of $A_{1g}$ mode indicate the weaker interlayer interactions in MoS₂ cannot be only associated with the van der Waals interactions, instead there are addition interlayer interactions in the material.[45] The $E_{2g}^1$ and the $A_{1g}$ vibrations in BNNT@multi-MoS₂NT both softened compared with these two modes in bulk MoS₂, which might reflect the tensile strain in nanotubes does have a strong effect on the phonon dispersion and causes softening of the two active Raman modes.[52]

Moreover, there are more Raman peaks in Figure 4c using the excitation wavelength of 632.8 nm which is in resonance with the direct band gap (~1.96 eV) of MoS₂ at the K point.[53] In the resonant Raman spectra, the most prominent mode around 460 cm⁻¹ arises from a second-order process of involving the longitudinal acoustic phonons at M point (LA(M)).[50] On the spectrum of BNNT@multi-MoS₂NT, the peaks around 178, 453, and 637 cm⁻¹ were observed as well the $E_{2g}^1$ (380 cm⁻¹) and $A_{1g}$ (405 cm⁻¹) peaks. The peak at 453 cm⁻¹ is assigned as the double frequency of the LA(M) mode. Moreover the peaks centered at 178 and 637 cm⁻¹ are assigned to $A_{1g} - $ LA and $A_{1g} + $ LA Raman modes, respectively. The shoulder of $A_{1g}$ peak in BNNT@multi-MoS₂NT evolves into an individual peak at 421 cm⁻¹ in BNNT@single-MoS₂NT. This peak has been interpreted through a Raman-inactive mode (B₁ᵤ) due to a two-photon scattering process involving a longitudinal quasi-acoustic phonon and a transverse optical phonon.[50] Moreover, on the Raman spectrum of BNNT@single-MoS₂NT, a blue-shift of the $E_{2g}^1$ (385 cm⁻¹) and $A_{1g}$ (407 cm⁻¹) peaks was both observed respect to the two modes in BNNT@multi-MoS$_2$NT. Interestingly, the intensity of $A_{1g} - \text{LA}$ and $A_{1g} + \text{LA}$ vibration modes become much weaker in BNNT@single-MoS$_2$NT. Besides, the resonant Raman spectrum of BNNT@single-MoS$_2$NT is gradually lift-up of background towards high frequencies which is attributed to the tail of the extremely strong PL in single-walled MoS$_2$ nanotubes. These significant differences of the Raman features between BNNT@single-MoS$_2$NT and BNNT@multi-MoS$_2$NT imply that Raman spectroscopy could serve as a reliable tool for identifying single-walled MoS$_2$ nanotubes.

The absorption spectrum of BNNT@single-MoS$_2$NT was found to be mostly unaltered, except for a slight blueshift of the resonances (Figure 4b). BNNT@single-MoS$_2$NT PL peak A at 1.85 eV almost matches the lower absorption resonance in its position (Figure 4d). The PL of BNNT@multi-MoS$_2$NT was significantly quenched, which indicates multi-walled MoS$_2$ nanotubes are similar to the bulk MoS$_2$ having an indirect band gap.

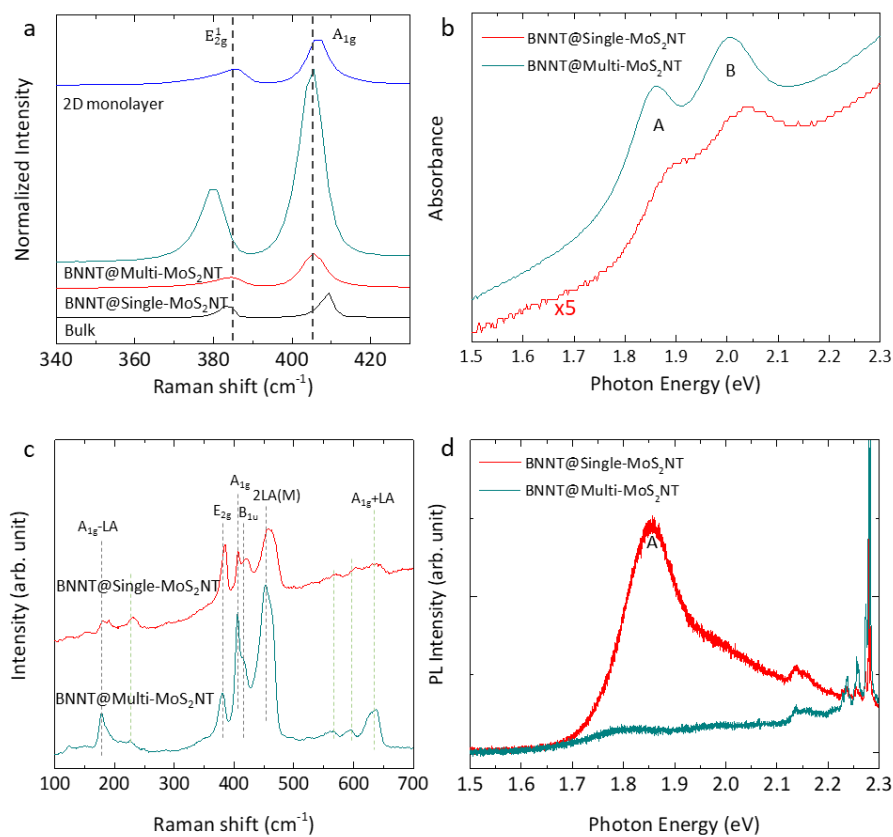

Figure 4. (a) Raman spectra of 2D monolayer MoS$_2$, BNNT@multi-MoS$_2$NT, BNNT@single-MoS$_2$NT, and the bulk MoS$_2$ with the excitation wavelength of 532 nm. (b) Representative absorption spectra of BNNT@single-MoS$_2$NT and BNNT@multi-MoS$_2$NT heteronanotubes. (c) Resonant Raman spectra of BNNT@single-MoS$_2$NT and BNNT@multi-MoS$_2$NT heteronanotubes using 632.8 nm laser. (d) PL spectra of BNNT@single-MoS$_2$NT and BNNT@multi-MoS$_2$NT heteronanotubes with 532 nm excited laser.

**Conclusions**

In conclusion, heteronanotubes of single- and multi-walled MoS$_2$ nanotubes coaxially grown on BNNTs were synthesized in this work. We observed a strong PL from single-walled MoS$_2$ nanotubes grown on BNNTs, and this result corresponded with our DFT results that large-diameter single-walled MoS$_2$ nanotubes have direct band gap at K point. Meanwhile, the PL signal of single-walled MoS$_2$ nanotubes in SWCNT@BNNT@MoS$_2$NT heteronanotubes is noticeably quenched. The charge and energy transfer between SWCNTs and single-walled MoS$_2$ nanotubes in one-dimensional van der Waals heterostructures were examined by PL and Raman spectroscopy. The significant PL quenching in multi-walled MoS$_2$ nanotubes verified their indirect band gap structures. The different Raman features of single- and multi-walled MoS$_2$ nanotubes under resonant and non-resonant condition may provide a tool for distinguishing the number of walls in MoS$_2$ nanotubes. The BNNT based hetero structure may pave a way for investigating the optical properties of TMD nanotubes.

## Methods

**Synthesis of the BNNT Film.** The SWCNT film used in this work as a template was synthesized by aerosol CVD and collected on a filter paper.[54] The SWCNT film with filter paper was dry-transferred onto a ceramic washer (6 mm). After removing the filter paper, the SWCNT film was suspended on the ceramic washer. For the BNNT CVD, the BN precursor, 30 mg of borane-ammonia complex (97%, Sigma-Aldrich), was placed upstream in the quartz tube and heated to 70°C. A low-pressure CVD with 300 Pa was applied here for the reaction at 1075°C and Ar with 3% $H_2$ was the carrier gas for the BN precursor with a flow rate of 300 sccm. The representative CVD time here was 1 hour and the average coating ratio was around 55%. After the BNNT coating CVD procedure, samples were set in a pure $O_2$ atmosphere CVD system for thermal oxidizing the inner SWCNT film to obtain the pure BNNT film. The oxidation process was taken at 610°C at 85kPa with a flow rate of 45 sccm $O_2$ gas for 12 hours.

**Growth of the BNNT@$MoS_2$NT film.** The BNNT@$MoS_2$NT heteronanotube was synthesized by a low-pressure CVD system using sulfur (sublimated, 99.0%, Wako) and $MoO_3$ (99.9%, Sigma-Aldrich) powders as precursors. Sulfur(S) powder was loaded in a quartz boat and placed out of the furnace at the upstream of the quartz tube chamber. The S quartz boat was heated up by a rubber heater to 138 °C. $MoO_3$ powder was placed upstream of the substrates and heated up to 530 °C. The BNNT film sample was set downstream of the $MoO_3$ powder and heated up to 530 °C for synthesizing single-walled $MoS_2$ nanotubes and 750 °C for multi-walled $MoS_2$ nanotubes. Ar gas with a flow rate of 50 sccm was introduced into the chamber as a carrier. The typical reaction time at aimed temperature was 10 min. The distance of the $MoO_3$ quartz boat and the substrate is an essential parameter for controlling the quality of $MoS_2$. The distance in this experiment was 10 cm.

**Optical characterizations and TEM characterizations.** The Raman and PL spectra of the samples were obtained by a Raman spectrometer (inVia, Renishaw) with the excitation wavelengths of 532 nm and 632.8 nm. The absorption spectra were measured by a UV-Vis-NIR spectrophotometer (Shimadzu UV-3150). The photo-emission measurements were performed using XPS (PHI5000, Versa Probe) with monochromatic Al Kα radiation. The TEM images were obtained by a JEM-2010F microscope and a JEM-2800 microscope at an acceleration voltage of

200 kV or 100 kV. The high-resolution TEM images were taken by a JEM-ARM200F microscope with a thermal field-emission gun operating at 120 kV.

**DFT Calculation.** The Simulation Tool for Atom Technology (STATE)[55] package was used for investigating the geometric and electronic structures. General gradient approximation by Perdew-Burke-Ernzerhof functional[56, 57] was applied to describe exchange correlation potential energy between electrons. Ultrasoft pseudopotentials were used for the pseudopotential between electrons and ions.[58] Cutoff energies of plane-wave to expand the valence wavefunction and deficit charge density were 340 and 3061 eV (25 and 225 Ry), respectively. Atomic structure was optimized until forces acting on each atom are less than 0.684 eV/nm ($1.33 \times 10^{-3}$ Hartree/au). A lattice parameter along tube direction was fixed to 0.315 nm, which corresponds with the experimental value of bulk $MoS_2$.[59] The Brillouin zone integration was carried out under six $k$ points along the tube direction.

**Acknowledgements**

**Part of this work was supported by JSPS KAKENHI Grant Numbers JP18H05329, JP19H02543, JP20H00220, JP20KK0114, JP18J22263 and by JST, CREST Grant Number JPMJCR20B5, Japan. ML acknowledges the support from JSPS Grant-in-Aid for Young Scientist Grant Number JP19J13441. Part of the work was conducted at the Advanced Characterization Nanotechnology Platform of the University of Tokyo, supported by the "Nanotechnology Platform" of the MEXT, Japan, grant numbers JPMXP09A20UT0063.**